\documentclass[a4paper]{article}

\usepackage[latin1]{inputenc}
\usepackage{amsfonts}
\usepackage{amsmath}
\usepackage{amssymb}
\usepackage[UKenglish]{babel}
\usepackage{fontenc}
\usepackage{graphicx}
\usepackage{amscd}
\usepackage{psfrag}
\usepackage{amscd}
\usepackage{amsthm}

\newcommand{\CC}{\ensuremath{\mathbb{C}}}
\newcommand{\RR}{\ensuremath{\mathbb{R}}}
\newcommand{\NN}{\ensuremath{\mathbb{N}}}
\newcommand{\ZZ}{\ensuremath{\mathbb{Z}}}

\newcommand{\PP}{\ensuremath{\mathbb{P}}}
\newcommand{\CP}[1]{\ensuremath{\mathbb{CP}^{#1}}}

\renewcommand{\d}{\ensuremath{\mathrm{d}}}
\newcommand{\covd}{\ensuremath{\nabla}}

\newcommand{\modsq}[1]{\ensuremath{\left|#1\right|^2}}

\newcommand{\Ob}[1]{\ensuremath{\mathcal{O}(#1)}} 		
\newcommand{\Os}{\ensuremath{\mathcal{O}}}			
\newcommand{\epsi}{\varepsilon}

\newcommand{\Aut}{\ensuremath{\mathrm{Aut}}}

\newcommand{\U}[1]{\ensuremath{\mathrm{U}(#1)}}

\newcommand{\im}{\ensuremath{\bold{i}}}
\newcommand{\half}{\ensuremath{\frac{1}{2}}}

\newcommand{\ihalf}{\ensuremath{\frac{\im}{2}}}

\newcommand{\w}{\wedge}

\newcommand{\e}{\ensuremath{\mathrm{e}}}

\renewcommand{\dim}[1]{\ensuremath{\mathrm{dim}(#1)}}

\newcommand{\fix}{\ensuremath{\mathrm{Fix}}}

\newcommand{\ga}{\gamma}
\newcommand{\de}{\delta}

\newcommand{\vol}[1]{\ensuremath{\mathrm{Vol}(#1)}}

\newcommand{\curlM}{\ensuremath{\mathcal{M}}}

\newcommand{\curlL}{\ensuremath{\mathcal{L}}}

\newcommand{\curlS}{\ensuremath{\mathcal{S}}}

\newcommand{\Hc}{\ensuremath{\mathrm{H}}}

\newcommand{\pd}{\ensuremath{\partial}}

\newcommand{\ph}{\phantom}
\newcommand{\ol}{\overline}

\newcommand{\Sym}{\ensuremath{\mathrm{Sym}}}
\newcommand{\AJ}{\ensuremath{\mathrm{AJ}}}
\newcommand{\Jac}{\ensuremath{\mathrm{Jac}}}
\newcommand{\Pic}{\ensuremath{\mathrm{Pic}}}
\newcommand{\spn}{\ensuremath{\mathrm{span}}}

\begin{document}

\pagestyle{plain}

\title{
\vskip -70pt
\begin{flushright}
{\normalsize DAMTP-2013-19} \\
\end{flushright}
\vskip 50pt
{\bf \Large Vortices and the Abel--Jacobi map}
\vskip 30pt
}

\author{
Norman A. Rink\footnote{nar43@cantab.net} \\ \\
{\sl Department of Applied Mathematics and Theoretical Physics,}\\
{\sl University of Cambridge,}\\
{\sl Wilberforce Road, Cambridge CB3 0WA, England.}\\
}

\vskip 20pt
\date{23 February 2014}
\vskip 20pt

\maketitle

\begin{abstract}

The abelian Higgs model on a compact Riemann surface $\Sigma$ supports vortex solutions for any positive vortex number $d\in\ZZ$. Moreover, the vortex moduli space for fixed $d$ has long been known to be the symmetrized $d$-th power of $\Sigma$, in symbols, $\Sym^d(\Sigma)$. This moduli space is K\"ahler with respect to the physically motivated metric whose geodesics describe slow vortex motion. 

In this paper we appeal to classical properties of $\Sym^d(\Sigma)$ to obtain new results for the moduli space metric. Our main tool is the Abel--Jacobi map, which maps $\Sym^d(\Sigma)$ into the Jacobian of $\Sigma$. Fibres of the Abel--Jacobi map are complex projective spaces, and the first theorem we prove states that near the Bradlow limit the moduli space metric restricted to these fibres is a multiple of the Fubini--Study metric. Additional significance is given to the fibres of the Abel--Jacobi map by our second result: we show that if $\Sigma$ is a hyperelliptic surface, there exist two special fibres which are geodesic submanifolds of the moduli space. Even more is true: the Abel--Jacobi map has a number of fibres which contain complex projective subspaces that are geodesic.

\end{abstract}

\newpage

\newpage

\section{Introduction and results} \label{sec:intro}

Vortices in the abelian Higgs model on a Riemann surface $\Sigma$ have been studied for a long time \cite{Nielsen:Olesen, Bogomolny, Witten, Jaffe:Taubes, Samols, Bradlow:line:bundles, GP:direct, Manton:Sutcliffe}, and the vortex moduli space has also received much attention. In the so-called geodesic approximation \cite{Manton:BPS:1982, Stuart:dynamics} the motion of vortices is described by geodesics on the moduli space, with respect to a physically motivated metric. The moduli space is K\"ahler with respect to this metric \cite{Samols, GP:direct}, and a semi-explicit, local formula for the moduli space metric was also obtained in \cite{Samols}. Moreover, one can give an explicit description of the moduli space as a complex manifold: this is due to the result in \cite{Bradlow:line:bundles} that vortex configurations on $\Sigma$, with vortex number $d$, are in 1-1 correspondence with positive divisors of degree $d$. Hence the moduli space is the symmetrized power $\Sym^d(\Sigma)$, and, rather intuitively, a positive divisor $D\in\Sym^d(\Sigma)$ describes a configuration of vortices centred at the points in the support of $D$.

Here we are solely interested in the case where $\Sigma$ is a compact Riemann surface. Then, viewing the moduli space as $\Sym^d(\Sigma)$, one can use classical methods from the theory on compact Riemann surfaces to gain further insight into the structure of the moduli space. In particular, there is a version of the Abel--Jacobi map,
\begin{align}
 \AJ \colon \Sym^d(\Sigma) \to \Jac(\Sigma), \label{eq:Abel:Jacobi}
\end{align}
which maps the vortex moduli space into the Jacobian $\Jac(\Sigma)$ of the Riemann surface $\Sigma$. The image of $\AJ$ is generally a subvariety of $\Jac(\Sigma)$, and the fibres are complex projective spaces. The Abel--Jacobi map is particularly useful when studying vortices near the so-called Bradlow limit \cite{Bradlow:line:bundles}. The Bradlow limit provides an upper bound on the number of vortices that can fit on a surface $\Sigma$ of finite area. In the Bradlow limit vortices are fully dissolved, i.e.~the Higgs field vanishes identically. Near the Bradlow limit the Higgs field magnitude is small and therefore vortices are very spread out objects. The smallness of the Higgs field magnitude suggests that one should expand  interesting quantities, such as the moduli space metric, in terms of the Higgs field magnitude. 

The usefulness of the Abel--Jacobi map near the Bradlow limit is a consequence of the fact that in the strict Bradlow limit the vortex moduli space coincides with $\Jac(\Sigma)$. The moduli space metric in the Bradlow limit was recently shown in \cite{Manton:Romao:Jac} to be the flat metric on the torus $\Jac(\Sigma)$. Near the Bradlow limit the fibres of $\AJ$ are metrically small, and the lowest order contribution to the moduli space metric is due to vortex motion in the image of $\AJ$. This contribution was also worked out in \cite{Manton:Romao:Jac}, and it was found to be degenerate at those ${\mathfrak p}\in\Jac(\Sigma)$ for which $\AJ^{-1}\{{\mathfrak p}\}$ has positive dimension. This leads to the question what contribution to the moduli space metric is due to vortex motion in the fibres of $\AJ$. The answer to this question is our first theorem.

\newtheorem{FS:metric}{Theorem}
\begin{FS:metric} \label{thm:FS:metric}
Let ${\mathfrak p}\in\Jac(\Sigma)$ be in the image of $\AJ$, and let $k\in\NN$ be such that $\AJ^{-1}\{{\mathfrak p}\}\cong\CP{k}$. Near the Bradlow limit the leading order contribution to the moduli space metric, restricted to $\AJ^{-1}\{{\mathfrak p}\}$, is a multiple of the Fubini--Study metric on $\CP{k}$.
\end{FS:metric}

This result generalizes the work of \cite{Baptista:Manton:S2}, where vortices on $\CP{1}$ were studied near the Bradlow limit, and the moduli space metric was found to be a multiple of the Fubini--Study metric. Note that $\Sym^d(\CP{1})\cong\CP{d}$, and since the Jacobian of $\CP{1}$ is a point, the Abel--Jacobi map is trivial. Hence, the result of \cite{Baptista:Manton:S2} is a special case of Theorem \ref{thm:FS:metric}.  

While Theorem \ref{thm:FS:metric} serves to alleviate the degeneracy of the moduli space metric found in \cite{Manton:Romao:Jac}, the splitting of vortex motion into directions in the image and fibre of $\AJ$ does not appear to be very natural from a physical point of view. However, if a fibre of $\AJ$ is a geodesic submanifold of the moduli space, then it is sensible to study the dynamics of vortex configurations that correspond to points in this fibre. Our second result is that on a hyperelliptic Riemann surface there exist two fibres of $\AJ$ that are indeed geodesic. 

Before we can express the previous statement as a theorem, we need to introduce some notation: by $K_{\Sigma}$ we denote the canonical line bundle of the Riemann surface $\Sigma$. A hyperelliptic Riemann surface $\Sigma$ is defined by the existence of a holomorphic projection map
\begin{align}
 \pi\colon\Sigma\to\CP{1},
\end{align}
which is 2-1. If we regard a point $p\in\CP{1}$ as a divisor of degree one, then the pulled-back divisor $\pi^{-1}(p)$ consists of the two points, counted with multiplicity, in the preimage of $\pi$. The holomorphic line bundle $\Ob{\pi^{-1}(p)}$ on $\Sigma$ is independent of $p$, up to isomorphism. The existence of the map $\pi$ also leads to a natural holomorphic automorphism of $\Sigma$, defined by exchanging the two points in $\pi^{-1}(p)$. This automorphism is referred to as the hyperelliptic involution of $\Sigma$. Now we are ready to state our next theorem.

\newtheorem{geodesic}[FS:metric]{Theorem}
\begin{geodesic} \label{thm:geodesic}
Let $\Sigma$ be a hyperelliptic Riemann surface, equipped with a metric such that the hyperelliptic involution of $\Sigma$ is an isometry. Then $\PP\Hc^0(\Sigma,K_{\Sigma})$ and $\PP\Hc^0(\Sigma,\Ob{\pi^{-1}(p)})$, $p\in\CP{1}$, are geodesic submanifolds of the moduli space.
\end{geodesic}

As usual, $\Hc^0(\Sigma,K_{\Sigma})$ denotes the space of holomorphic sections of $K_{\Sigma}$, and $\PP\Hc^0(\Sigma,K_{\Sigma})$ is its projectivization, and analogously for $\Ob{\pi^{-1}(p)}$. It is a standard observation that $\PP\Hc^0(\Sigma,K_{\Sigma})$ and $\PP\Hc^0(\Sigma,\Ob{\pi^{-1}(p)})$ can be identified with fibres of the Abel--Jacobi map, and we will review this in section \ref{sec:line:bundles:AJ}. We stress that Theorem \ref{thm:geodesic} holds in general, not only near the Bradlow limit. The fact that the fibre $\PP\Hc^0(\Sigma,\Ob{\pi^{-1}(p)})$ is a geodesic submanifold of the vortex moduli space is also a consequence of Proposition 7.1 in \cite{Bockstedt:Romao}, for which an isometry on the moduli space is required.

Theorem \ref{thm:geodesic} is a special case of Lemma \ref{thm:geodesic:decomposition}, which we will establish in section \ref{sec:abelian:hyperelliptic}. In this introduction we omit a precise statement of Lemma \ref{thm:geodesic:decomposition} since this requires some additional technical preparations. In summary, Lemma \ref{thm:geodesic:decomposition} identifies a number of complex projective spaces that embed into the vortex moduli space as geodesic submanifolds. These complex projective spaces are linear systems of divisors on $\Sigma$. However, these linear systems are not necessarily complete, i.e.~they may not fill out entire fibres of the Abel--Jacobi map.

The structure of this paper is as follows. In section \ref{sec:line:bundles:AJ} we review classification results for holomorphic line bundles on compact Riemann surfaces. This review includes the definition of the Abel--Jacobi map and general properties of its fibres. In section \ref{sec:abelian:metric} we study the Bogomolny equations near the Bradlow limit. We introduce a suitable description of the moduli space near the Bradlow limit, and based on this we prove Theorem \ref{thm:FS:metric}. In section \ref{sec:abelian:hyperelliptic} we review standard properties of hyperelliptic Riemann surfaces and we use them to prove Lemma \ref{thm:geodesic:decomposition}, and consequently Theorem \ref{thm:geodesic}.

\section{Holomorphic line bundles on Riemann surfaces} \label{sec:line:bundles:AJ}

In this section we summarize the standard classification results for holomorphic line bundles in terms of the Picard and Jacobian varieties. We also review how holomorphic line bundles can be characterized equivalently by divisors and by Dolbeault operators. For detailed derivations and proofs we refer to standard textbooks such as \cite{Farkas:Kra, Forster, Jost, Gunning:RS}.  The main purpose of this section is to introduce notation, and the reader familiar with the theory of line bundles on compact Riemann surfaces may well want to skip to section \ref{sec:abelian:metric}.

\subsection{Classification of holomorphic line bundles}

Isomorphism classes of holomorphic line bundles are in 1-1 correspondence with elements of the cohomology group $\Hc^1(\Sigma,\Os^*)$, where $\Os^*$ denotes the sheaf of nowhere vanishing holomorphic functions on $\Sigma$. Topologically a line bundle on $\Sigma$ is fully classified by its degree, but the classification of holomorphic line bundles is finer than this.\footnote{Note that in one complex dimension the degree of a line bundle agrees with its first Chern class.} 

The exponential sequence,
\begin{align}
 0 \to \ZZ \to \Os \to \Os^* \to 0,
\end{align}
leads to the classification of holomorphic line bundles which are topologically trivial. In cohomology we obtain the long exact sequence,
\begin{align}
 0 \to \Hc^1(\Sigma,\ZZ) \to \Hc^1(\Sigma,\Os) \to \Hc^1(\Sigma,\Os^*) \to \Hc^2(\Sigma,\ZZ) \to 0, \label{eq:cohomology:sequence}
\end{align}
where the last non-trivial map is the degree of $\curlL\in\Hc^1(\Sigma,\Os^*)$. The kernel of this map is isomorphic to the Picard variety,
\begin{align}
 \Pic_0(\Sigma) = \Hc^1(\Sigma,\Os)/\Hc^1(\Sigma,\ZZ) \cong \CC^g/\ZZ^{2g}, 
\end{align}
where $g$ is the genus of $\Sigma$. It follows that holomorphic line bundles of fixed degree $d$ are classified by points in $\Pic_0(\Sigma)$, up to isomorphism.

An alternative classification of holomorphic line bundles is in terms of the Jacobian of $\Sigma$. Let $\ga_k$, $k=1,\dots,2g$, be a basis of $\Hc_1(\Sigma,\ZZ)\cong\ZZ^{2g}$, and let $\omega_1, \dots, \omega_g$ be a basis of holomorphic 1-forms on $\Sigma$. Then the $2g$ vectors
\begin{align}
 \left( \int_{\ga_k}\omega_1, \dots , \int_{\ga_k}\omega_g \right), \quad k=1,\dots,2g, 
\end{align}
span a lattice $\Lambda\cong\ZZ^{2g}$. The Jacobian of $\Sigma$ is defined as the torus 
\begin{align}
 \Jac(\Sigma) = \CC^g/\Lambda,
\end{align}
and
\begin{align}
 \Jac(\Sigma) \cong \Pic_0(\Sigma),
\end{align}
proofs of which can be found in \cite{Forster, Gunning:RS}.

\subsection{Divisors and the Abel--Jacobi map} \label{sec:line:bundles:divisors:AJ:map}

We now present a version of the Abel--Jacobi map which is useful in the context of vortices: the $d$-th symmetrized power of $\Sigma$, denoted as $\Sym^d(\Sigma)$, can be regarded as the space of positive divisors of degree $d$. To define the Abel--Jacobi map, 
\begin{align}
  &\AJ\colon \Sym^d(\Sigma) \to \Jac(\Sigma), \label{eq:sym:AJ}
\end{align}
choose a fixed divisor ${\tilde D}=\sum_{l=1}^d {\tilde x}_l \in\Sym^d(\Sigma)$, and set
\begin{align}
  &\AJ(D) = \sum_{l=1}^d \left( \int_{{\tilde x}_l}^{x_l}\omega_1, \dots , \int_{{\tilde x}_l}^{x_l}\omega_g \right) \,{\rm mod}\, \Lambda \:,	
\end{align}
for $D = \sum_{l=1}^d x_l \in \Sym^d(\Sigma)$. To identify the fibres of the map $\AJ$, take two positive divisors  $D, D'$. By Abel's Theorem (cf.~\cite{Forster}) $\AJ(D) = \AJ(D')$ holds if and only if $D$ and $D'$ differ by a principal divisor, i.e.~if and only if there exists a meromorphic function $f\colon\Sigma\to\CC$ such that $D'=D+(f)$, where $(f)$ denotes the divisor of $f$. If we let
\begin{align}
 {\mathfrak L}(D) = \{ f\colon\Sigma\to\CC \colon f\text{ meromorphic and } (f)\ge-D \},
\end{align}
then the preimage of $\AJ(D)$ is simply given by
\begin{align}
 \AJ^{-1}\{{\rm AJ}(D)\} = \PP{\mathfrak L}(D). \label{eq:AJ:inverse:image}
\end{align}
Here $\PP{\mathfrak L}(D)$ is the projectivization of ${\mathfrak L}(D)$. A collection of divisors that form a (projective) linear subspace of $\PP{\mathfrak L}(D)$ is called a linear system, and $\PP{\mathfrak L}(D)$ is referred to as the complete linear system of $D$. Identity \eqref{eq:AJ:inverse:image} holds because the quotient $f'/f$ of two meromorphic functions $f$, $f'$ which satisfy $(f)=(f')$ is necessarily a constant function. Expressed in words, identity \eqref{eq:AJ:inverse:image} says that the fibres of $\AJ$ are precisely the complete linear systems of divisors on $\Sigma$.   

If $D\in\Sym^d(\Sigma)$, then the line bundle $\Ob{D}$ has a non-trivial holomorphic section. Let $\phi_D$ be such a section, with divisor $(\phi)=D$, and denote as $\Hc^0(\Sigma,\Ob{D})$ the space of all holomorphic sections of $\Ob{D}$. The following map is an isomorphism,
\begin{align}
 {\mathfrak L}(D) &\to \Hc^0(\Sigma,\Ob{D}) \\
 f &\mapsto f\phi_D,
\end{align}
and hence yields another interpretation of the preimage of $\AJ(D)$,
\begin{align}
 \AJ^{-1}\{{\rm AJ}(D)\} \cong \PP\Hc^0(\Sigma,\Ob{D}).
\end{align}

\subsection{Dolbeault operators and automorphisms} \label{subsec:Dolbeault}

An alternative but equivalent way to describe holomorphic line bundles is in terms of Dolbeault operators. If $L$ is a smooth line bundle over $\Sigma$, with smooth sections $\Gamma(\Sigma,L)$, then a Dolbeault operator is a derivative operator
\begin{align}
 {\covd}^{0,1} \colon \Gamma(\Sigma,L) \to \Omega^{0,1}(\Sigma,L),
\end{align}
which satisfies the Leibniz rule and
\begin{align}
 {\covd}^{0,1} \circ {\covd}^{0,1} = 0.
\end{align}
By $\Omega^{0,1}(\Sigma,L)$ we denote the $(0,1)$-form valued sections of $L$. The pair $(L,{\covd}^{0,1})$ is a holomorphic line bundle since, by the Leibniz rule, the transition functions defining $(L,{\covd}^{0,1})$ are necessarily holomorphic.

We denote as $\Aut(L)$ the space of smooth automorphisms of the line bundle $L$, i.e.
\begin{align}
 \Aut(L) = \{g\colon \Sigma \to \CC^* \text{ smooth}\}.
\end{align}
Two Dolbeault operators $\covd^{0,1}$ and ${\tilde\covd}^{0,1}$ on $L$ yield isomorphic holomorphic line bundles if they are related by 
\begin{align}
 {\tilde\covd}^{0,1} = g\circ \covd^{0,1}\circ g^{-1},
\end{align}
where $g\in\Aut(L)$. Denoting the space of Dolbeault operators on $L$ as ${\mathfrak A}^{0,1}(L)$, it follows that there is a 1-1 correspondence
\begin{align}
 {\mathfrak A}^{0,1}(L) / \Aut(L) \,\leftrightarrow\, \Jac(\Sigma),
\end{align}
where the quotient on the left-hand side is the space of $\Aut(L)$-orbits in ${\mathfrak A}^{0,1}(L)$. 

We now assume that the smooth line bundle $L$ carries a hermitian structure, and we let $\covd$ be a compatible covariant derivative. The covariant derivative can be decomposed into its $(1,0)$ and $(0,1)$ parts,
\begin{align}
 \covd = \nabla^{1,0} + \covd^{0,1},
\end{align}
and the $(0,1)$-part $\covd^{0,1}$ is a Dolbeault operator since $\covd^{0,1}\circ\covd^{0,1} = 0$ on dimensional grounds. We choose to work in unitary gauge, in which we write locally
\begin{align}
 \covd = \d + A, \quad \covd^{1,0} = \pd + A^{1,0}, \quad \covd^{0,1} = \bar\pd + A^{0,1}, 
\end{align}
with a 1-form $A = A^{1,0} + A^{0,1}$. Since $\covd$ is compatible with the hermitian structure on $L$, we have that $A$ is anti-hermitian, i.e.~${\bar A} = -A$. The $\Aut(L)$-action on $\covd^{0,1}$ can be expressed as
\begin{align}
 A^{0,1} \mapsto A^{0,1} - (\bar\pd g)g^{-1}, 
\end{align}
where $g\in\Aut(L)$. Note that this action extends to the full covariant derivative,
\begin{align}
 g(\nabla) = \d + A + (\pd\bar g){\bar g}^{-1} - (\bar\pd g)g^{-1}. \label{eq:g:action}
\end{align}
The advantage of this action is that $g(\covd)$ is again compatible with the same hermitian structure (cf.~\cite{Bradlow:line:bundles} or section VII.1 in \cite{Kobayashi:vector:bundles}). It should be noted that $g$ is a unitary gauge transformation precisely if $g\colon\Sigma\to\U{1}$, and in this case \eqref{eq:g:action} defines a unitary gauge transformation of the gauge potential $A$. Another observation worth making is that the curvature $F_{\covd} = \covd\circ\covd$ is related to $F_{g(\covd)}$ by
\begin{align}
 F_{g(\covd)} = F_{\covd} + \bar\pd \pd \log\modsq{g}. \label{eq:curvature:action}
\end{align}
If $g$ is a unitary gauge transformation, then of course $\modsq{g}=1$, and hence $F_{g(\covd)} = F_{\covd}$.

\section{The vortex moduli space} \label{sec:abelian:metric}

This section contains the proof of Theorem \ref{thm:FS:metric}. A few preparations are necessary to set up in a precise way the Bogomolny equations near the Bradlow limit. We begin by reviewing the Bogomolny equations for abelian vortices on $\Sigma$ and their solution theory. 

Let $z$ be a local complex coordinate on the Riemann surface $\Sigma$. The metric and volume form on $\Sigma$ are given by
\begin{align}
 \d{s}^2 = \Omega_{\Sigma}(z,\bar z) \d{z}\d{\bar z}, \quad \omega_{\Sigma} = \ihalf \Omega_{\Sigma}(z,\bar z) \d{z}\w\d{\bar z}.
\end{align}
As before, let $L$ be a smooth line bundle over $\Sigma$, equipped with a hermitian structure. We continue to work in unitary gauge. If the degree of $L$ is $d>0$, then a configuration of $d$ vortices in the abelian Higgs model is a solution of the Bogomolny equations
\begin{align}
 &{\covd}_{\bar z}\phi = 0,  \label{eq:ab:Bog:1} \\
 &F_{z\bar z} = \frac{\Omega_{\Sigma}}{4} \left(1 - \modsq{\phi} \right), \label{eq:ab:Bog:2}
\end{align}
where $\phi$ is a section of $L$, $\covd$ is a covariant derivative compatible with the hermitian structure, and $F = \covd\circ\covd$ is the curvature of $\covd$. Note that $\covd^{0,1} = \d{\bar z}\,{\covd}_{\bar z}$, and hence \eqref{eq:ab:Bog:1} says that $\phi$ is a holomorphic section of $(L,\covd^{0,1})$. As in the previous section we write $\covd = \d + A$, with the local gauge potential $A$ such that ${\bar A} = -A$. 

By integrating \eqref{eq:ab:Bog:2} one sees that the Bogomolny equations  have solutions only if 
\begin{align}
 4\pi d \le \vol{\Sigma}. \label{eq:Bradlow}
\end{align}
The situation where $4\pi d = \vol{\Sigma}$ is referred to as the Bradlow limit \cite{Bradlow:line:bundles}. In the Bradlow limit $\phi = 0$ identically, and hence the Bogomolny equations reduce to
\begin{align}
 &F_{z\bar z} = \frac{\Omega_{\Sigma}}{4}.  \label{eq:Bog:Bradlow:limit}
\end{align}
It was shown in \cite{Bradlow:line:bundles}, and again more recently in \cite{Manton:Romao:Jac}, that solutions of \eqref{eq:Bog:Bradlow:limit}, modulo unitary gauge transformations, are in 1-1 correspondence with points in $\Jac(\Sigma)$. This can also be expressed as follows, using the results summarized in section \ref{sec:line:bundles:AJ}. 

\newtheorem{Bradlow:limit}[FS:metric]{Theorem}
\begin{Bradlow:limit}[due to \cite{Bradlow:line:bundles, Manton:Romao:Jac}] \label{thm:Bradlow:limit}
Let $\curlL$ be a holomorphic line bundle on $\Sigma$ with underlying smooth line bundle $L$. Let $L$ be equipped with a hermitian structure, and let the degree $d$ of $L$ be such that equality holds in \eqref{eq:Bradlow}. Then there exists a Dolbeault operator $\covd^{0,1}$ on $L$, unique up to unitary gauge transformations, such that $\curlL \cong (L, \covd^{0,1})$ and the curvature of the compatible covariant derivative $\covd =\covd^{1,0} + \covd^{0,1}$ satisfies \eqref{eq:Bog:Bradlow:limit}.
\end{Bradlow:limit}

Note that the covariant derivative $\covd$ in the theorem is fully determined by the Dolbeault operator $\covd^{0,1}$ and the requirement that it be compatible with the hermitian structure on $L$.

Away from the Bradlow limit the moduli space of solutions of \eqref{eq:ab:Bog:1}, \eqref{eq:ab:Bog:2} also has a simple description: 
provided \eqref{eq:Bradlow} is a genuine inequality, solutions of the Bogomolny equations, modulo unitary gauge transformations, are in 1-1 correspondence with positive divisors of degree $d$. This was also shown in \cite{Bradlow:line:bundles}, and again later in \cite{GP:direct}. We also phrase this as a theorem about Dolbeault operators.

\newtheorem{moduli:space}[FS:metric]{Theorem}
\begin{moduli:space} [due to \cite{Bradlow:line:bundles, GP:direct}] \label{thm:moduli:space}
For every divisor $D\in\Sym^d(D)$ there exists a Dolbeault operator $\covd^{0,1}$ on $L$ and $\phi\in\Gamma(\Sigma, L)$ such that
\begin{enumerate}
 \item $(L,\covd^{0,1}) = \Ob{D}$, $\phi\in\Hc^0(\Sigma,\Ob{D})$.
 \item $D$ is the divisor of zeros of $\phi$, counted with multiplicities, i.e.~$(\phi)=D$.
 \item The compatible covariant derivative $\covd =\covd^{1,0} + \covd^{0,1}$ satisfies \eqref{eq:ab:Bog:2}.
\end{enumerate}
The pair $(\covd,\phi)$ is uniquely determined by $D$, up to unitary gauge transformations.
\end{moduli:space}

In the next subsection we define the problem of solving the Bogomolny equations near the Bradlow limit. We will establish a description of the moduli space in terms of the image and fibres of the Abel--Jacobi map.

\subsection{The Bogomolny problem near the Bradlow limit}

Assume that $\Omega_{\Sigma}$ is the conformal factor of $\Sigma$ in the Bradlow limit, i.e.
\begin{align}
  4\pi d = \ihalf \int_{\Sigma} \Omega_{\Sigma}(z,\bar z) \d{z}\w\d{\bar z}.
\end{align}
To move away from the Bradlow limit, we replace $\Omega_{\Sigma}$ with
\begin{align}
 \Omega_{\Sigma,\epsi} = (1+\epsi)\Omega_{\Sigma},
\end{align}
where $\epsi\ll 1$ is a dimensionless parameter. Hence the new volume form on $\Sigma$ is
\begin{align}
 \omega_{\Sigma,\epsi} = \ihalf \Omega_{\Sigma,\epsi} \d{z}\w\d{\bar z}.
\end{align}
Let $\covd = \d + A_0$ be a solution of the Bogomolny equations for $\epsi=0$, i.e.~a solution of \eqref{eq:Bog:Bradlow:limit}, with gauge potential $A_0$. To solve the Bogomolny equations with positive $\epsi$, we make the ansatz $\covd + \epsi A_1$ for the covariant derivative, where $A_1$ is a globally defined 1-form on $\Sigma$.\footnote{Note that since we are working in unitary gauge, $A_1$ must be purely imaginary in order for $\covd + \epsi A_1$ to be compatible with the hermitian structure on $L$.} Then the Bogomolny equations read
\begin{align}
 &\covd^{0,1}\phi + \epsi A_1^{0,1}\phi = 0, \\
 &\im F + \im \epsi \d A_1 = \frac{\omega_{\Sigma,\epsi}}{2} \left(1 - \modsq{\phi} \right),
\end{align}
where $F = \d A_0$. Consistency in the limit $\epsi\to 0$ requires that $\phi = \sqrt{\epsi}\, \phi_{\half}$, to leading order in $\epsi$. Hence the Bogomolny equations reduce to
\begin{align}
 &\covd^{0,1}\phi_{\half} = 0, \label{eq:Bog:leading:1} \\
 &\im \d A_1 = \frac{\omega_{\Sigma}}{2} \left(1 - \vert\phi_{\half}\vert^2 \right), \label{eq:Bog:leading:2}
\end{align}
to leading order in $\epsi$. We refer to \eqref{eq:Bog:leading:1}, \eqref{eq:Bog:leading:2} as the Bogomolny problem near the Bradlow limit. 

The major simplification in comparison to \eqref{eq:ab:Bog:1}, \eqref{eq:ab:Bog:2} is that the Dolbeault operator $\covd^{0,1}$ does not appear in \eqref{eq:Bog:leading:2}, whereas it enters \eqref{eq:ab:Bog:2} implicitly through the curvature. Thus, in order to solve the Bogomolny problem near the Bradlow limit, one first chooses a holomorphic line bundle $\curlL = (L,\covd^{0,1})$, where $\covd^{0,1}$ solves $\eqref{eq:Bog:Bradlow:limit}$, and such that $\Hc^0(\Sigma,\curlL)\ne\{0\}$. Then one chooses $\phi_{\half}\in\Hc^0(\Sigma,\curlL)$ such that it satisfies the constraint 
\begin{align}
 4\pi d = \int \vert\phi_{\half}\vert^2\omega_{\Sigma}. \label{eq:Bradlow:Higgs:constraint}
\end{align}
It follows from the Hodge decomposition of 2-forms on $\Sigma$ that \eqref{eq:Bog:leading:2} can be solved for $A_1$ if and only if \eqref{eq:Bradlow:Higgs:constraint} holds. Recalling from Theorem \ref{thm:Bradlow:limit} that points in the Jacobian are in 1-1 correspondence with solutions of \eqref{eq:Bog:Bradlow:limit}, we can give the following description of the moduli space of solutions of the Bogomolny problem near the Bradlow limit. 

\newtheorem{near:Bradlow:limit}{Lemma}
\begin{near:Bradlow:limit} \label{thm:near:Bradlow:limit}
Let $\Aut_{\U{1}}(L)$ be the group of unitary automorphisms of $L$, and let
\begin{align}
 \curlS = \{ ({\mathfrak p},\phi_{\half}) \colon {\mathfrak p}=\AJ(D) \text{, and $\phi_{\half}\in\Hc^0(\Sigma,\Ob{D})$ satisfies \eqref{eq:Bradlow:Higgs:constraint}} \}.
\end{align}
The group $\Aut_{\U{1}}(L)$ acts on $\curlS$ by 
\begin{align}
 ({\mathfrak p},\phi_{\half}) \mapsto ({\mathfrak p},g\phi_{\half}), \quad g\in\Aut_{\U{1}}(L).
\end{align}
Solutions of the Bogomolny problem near the Bradlow limit, modulo unitary gauge transformations, are in 1-1 correspondence with points in the orbit space $\curlS/\Aut_{\U{1}}(L)$.
\end{near:Bradlow:limit}

Note that in the $\Aut_{\U{1}}(L)$-action on $\curlS$ the transformation
\begin{align}
 \covd^{0,1} \mapsto g\circ\covd^{0,1}\circ g^{-1}, \quad g\in\Aut_{\U{1}}(L),
\end{align}
is implicit since, by subsection \ref{subsec:Dolbeault}, this does not change ${\mathfrak p}=\AJ(D)$, where $(L,\covd^{0,1})\cong\Ob{D}$. We denote the moduli space near the Bradlow limit as $\curlM = \curlS/\Aut_{\U{1}}(L)$. It is straightforward to verify that there is a bijection between $\curlM$ and $\Sym^d(\Sigma)$. Hence Lemma \ref{thm:near:Bradlow:limit} does not yield a moduli space different from the one in Theorem \ref{thm:moduli:space}. However, Lemma \ref{thm:near:Bradlow:limit} allows us to identify the subspaces in $\curlM$ that correspond to fibres of the Abel--Jacobi map: since ${\mathfrak p}=\AJ(D)$ is left invariant by $\Aut_{\U{1}}(L)$, it is meaningful to consider the restriction $\curlM\vert_{\mathfrak p}$. As before, we let $\covd = \d + A_0$ be the covariant derivative on $L$ that is associated with ${\mathfrak p}$ by virtue of Theorem \ref{thm:Bradlow:limit}. We assume that the gauge potential $A_0$ has been gauge fixed. This can be achieved, for example, by imposing Gauss' law (cf.~\cite{Samols, Manton:Sutcliffe, Manton:Romao:Jac}). Then the residual gauge group consists of those $g\in\Aut_{\U{1}}(L)$ that leave $\covd = \d + A_0$ invariant, i.e.~constant $g\in\U{1}$. If $\Hc^0(\Sigma,\Ob{D}) \cong \CC^{k+1}$, then \eqref{eq:Bradlow:Higgs:constraint} implies that $\phi_{\half}$ lies on a sphere $S^{2k+1}\subset \Hc^0(\Sigma,\Ob{D})$. Therefore,
\begin{align}
 \curlM\vert_{\mathfrak p} \cong S^{2k+1} / \U{1} \cong \PP\Hc^0(\Sigma,\Ob{D}) \cong \AJ^{-1}\{{\mathfrak p}\}.
\end{align}

\subsection{The moduli space metric near the Bradlow limit}

The vortex moduli space carries a physically relevant metric which is induced by the kinetic energy,
\begin{align}
 T = \half \int_{\Sigma} \dot\phi \,\dot{\bar\phi}\,\omega_{\Sigma,\epsi} - \half \int_{\Sigma} \dot{A}\w *\dot{A}. \label{eq:kinetic:energy}
\end{align}
The physical significance of this metric is that its geodesics describe the motion of vortices at low velocities \cite{Manton:BPS:1982, Stuart:dynamics}. Using the ansatz $A = A_0 + \epsi A_1$ for the gauge potential near the Bradlow limit, we obtain for the kinetic energy to linear order in $\epsi$,
\begin{align}
 T = \frac{\epsi}{2} \int_{\Sigma} \dot\phi_{\half}\, \dot{\bar\phi}_{\half}\,\omega_{\Sigma} - \half \int_{\Sigma} \dot{A}_0 \w *\dot{A}_0 - \epsi \int_{\Sigma} \dot{A}_0 \w *\dot{A}_1. \label{eq:dissolving:kinetic:energy}
\end{align}
It is clear that in the Bradlow limit, when $\epsi=0$, only the middle term is present. In this case $T$ induces a metric on $\Jac(\Sigma)$, which was derived in \cite{Manton:Romao:Jac}. The moduli space metric induced by the middle term in \eqref{eq:dissolving:kinetic:energy} near the Bradlow limit was also worked out in \cite{Manton:Romao:Jac}. This required an analysis different from the one for $\epsi=0$ since for $\epsi>0$ only a subvariety of $\Jac(\Sigma)$ is accessible, namely the image of $\AJ$. 

In order to prove Theorem \ref{thm:FS:metric}, we consider a fixed ${\mathfrak p}=\AJ(D)$ and the restriction of $T$ to $\curlM\vert_{\mathfrak p}$. The crucial observation in our argument is that $\dot{A}_0=0$. This is because ${\mathfrak p}$ determines $\covd = \d + A_0$, up to a unitary gauge transformation, and we can assume, as before, that $A_0$ has been brought into a fixed gauge by imposing Gauss' law. Thus the moduli space metric restricted to $\curlM\vert_{\mathfrak p}$ is induced by the kinetic energy
\begin{align}
 T = \frac{\epsi}{2} \int_{\Sigma} \dot\phi_{\half}\, \dot{\bar\phi}_{\half}\,\omega_{\Sigma}.
\end{align}

Again because of $\dot{A}_0=0$ the linearization of \eqref{eq:Bog:leading:1} reads
\begin{align}
 \covd^{0,1}\dot{\phi}_\half = 0.
\end{align}
Hence $\dot{\phi}_\half\in\Hc^0(\Sigma,\Ob{D})$. We now introduce a basis of $\Hc^0(\Sigma,\Ob{D})$,
\begin{align}
 \Hc^0(\Sigma,\Ob{D}) = \spn_{\CC}\{\sigma_i\colon i=0,\dots,k\}.
\end{align}
We therefore have the expansions
\begin{align}
 \phi_{\half} = \sum_{i=0}^{k} a_i\sigma_i, \quad
 \dot\phi_{\half} = \sum_{i=0}^{k} \dot{a}_i\sigma_i,
\end{align}
where the $a_i\in\CC$ are constant across $\Sigma$ but depend on time. The $a_i$ can be regarded as homogeneous coordinates on $\curlM\vert_{\mathfrak p} \cong \CP{k}$. Hence, up to a redundant overall factor, the $a_i$ are the vortex moduli along the fibre $\AJ^{-1}\{{\mathfrak p}\}$.

Letting $H_{ij} = \int_{\Sigma} \sigma_i \bar{\sigma}_j \, \omega_{\Sigma}$ for $i,j=0,\dots,k$, we obtain a hermitian metric on $\Hc^0(\Sigma,\Ob{D})$,
\begin{align}
 H(b,c) = \sum_{i,j=0}^{k} b_i H_{ij} \bar{c}_j,
\end{align}
for vectors $b = (b_0,\dots,b_k)$ and $c = (c_0,\dots,c_k)$, corresponding to holomorphic sections $\sum_{i=0}^k b_i\sigma_i$, $\sum_{i=0}^k c_i\sigma_i$. Then, with $a = (a_0,\dots,a_k)$,
\begin{align}
 T = \frac{\epsi}{2} H(\dot{a}, \dot{a}), \label{eq:kinetic:H}
\end{align}
and the constraint \eqref{eq:Bradlow:Higgs:constraint} reads
\begin{align}
 4\pi  d = H(a,a).
\end{align}
In order to turn \eqref{eq:kinetic:H} into the moduli space metric restricted to $\curlM\vert_{\mathfrak p}$, we need to fix the residual gauge freedom
\begin{align}
 a \mapsto \e^{\im\chi} a, 
\end{align}
where $\chi\in\RR$. Note that this $\U{1}$-action generates the tangent vector
\begin{align}
 \frac{\d}{\d\chi}\!\left(\e^{\im\chi}a\right)_{\chi=0} = \im a
\end{align}
at $\phi_{\half}\in\Hc^0(\Sigma,\Ob{D})$. Therefore the gauge fixed version of the tangent vector $\dot{a}$ is
\begin{align}
 \dot{a}_{\rm g.f.} 
 &= \dot{a} - \frac{H( \dot{a}, \im a )}{H( \im a, \im a )} \im a \\
 &= \dot{a} - \frac{H( \dot{a}, a )}{4\pi d} a,
\end{align}
which leads to the moduli space metric
\begin{align}
 H\vert_{\curlM\vert_p}
 = \frac{\epsi}{2} H(\dot{a}_{\rm g.f.} , \dot{a}_{\rm g.f.} ) 
 = \frac{\epsi}{2} \left( H(\dot{a},\dot{a}) - \frac{ H( \dot{a}, a ) \ol{H( \dot{a}, a ) } }{4\pi d} \right).
\end{align}
This can be brought into the standard form of the Fubini--Study metric, up to a constant overall factor, by choosing coordinates in which $H_{ij} = \de_{ij}$ and defining $a = a' \sqrt{4\pi d}$. We have thus established Theorem \ref{thm:FS:metric}.

\section{Vortices on hyperelliptic surfaces} \label{sec:abelian:hyperelliptic}

From now on we let $\Sigma$ be a compact hyperelliptic Riemann surface. The goal of this section is to prove Theorem \ref{thm:geodesic}. In the first subsection we review the definition of hyperelliptic surfaces and we collect some of their properties. We then proceed to establishing Lemma \ref{thm:geodesic:decomposition}, from which we will subsequently deduce Theorem \ref{thm:geodesic}.

\subsection{Properties of hyperelliptic surfaces} \label{sec:hyperelliptic:surfaces}

We think of the hyperelliptic Riemann surface $\Sigma$ of genus $g$ as an algebraic curve given by the equation
\begin{align}
 y^2 = F(x),
\end{align}
where $F$ is a polynomial of degree $2g+1$ or $2g+2$. Here $x$ and $y$ should be thought of as coordinates on independent Riemann spheres $\CP{1}$. Then the defining projection map of $\Sigma$ is
\begin{align}
 \pi \colon \Sigma &\to \CP{1} \\
 (x,y) &\mapsto x.  \label{eq:hyperelliptic:covering}
\end{align}
By the Riemann--Hurwitz formula the map $\pi$ has $2g+2$ ramification points. The ramification points are pairs $(x,0)$, where $x$ is a root of the polynomial $F$. If $F$ has degree $2g+1$, then one ramification point is at $x=\infty$. We can redefine $x$ by acting on the $\CP{1}$ in the image of $\pi$ with a M\"obius transformation. A suitable M\"obius transformation will achieve that there is no ramification point at $x=\infty$. If $y$ is transformed accordingly, then we obtain again the defining equation of $\Sigma$ in the form $y^2 = F(x)$, but this time $F$ has degree $2g+2$. Without loss of generality we therefore assume that the degree of $F$ is $2g+2$ and that there is no ramification point at $x=\infty$. In a similar fashion we can also ensure that there is no ramification point at $x=0$.

Abstractly we denote the ramification points of $\pi$ as $r_1,\dots,r_{2g+2}\in\Sigma$. We also introduce the points $s_1, s_2, t_1, t_2 \in\Sigma$ which are defined by
\begin{align}
 &\pi(s_1) = \pi(s_2) = \infty, \\
 &\pi(t_1) = \pi(t_2) = 0.
\end{align}
Since there are no ramification points at $x=\infty$ or $x=0$, we have that $s_1\ne s_2$ and $t_1\ne t_2$. We can think of the map $\pi \colon \Sigma \to \CP{1}$ as a meromorphic function on $\Sigma$. If we adopt this point of view, we denote this meromorphic function also as $x$. Similarly we have a meromorphic function $y$ on $\Sigma$, which is given by the projection of $(x,y)$ onto the second component. Using the previously defined points on $\Sigma$, we can write down explicitly the divisors of $x$ and $y$, which are of degree zero,
\begin{align}
 &(x) = t_1 + t_2 - s_1 - s_2, 				\label{eq:x:divisor} \\
 &(y) = r_1 + \dots + r_{2g+2} - (g+1)s_1 - (g+1)s_2.	\label{eq:y:divisor}
\end{align}
The second line is a consequence of the equation $y^2=F(x)$. Since $x$ is a meromorphic function, $\d{x}$ is a meromorphic 1-form, whose divisor is
\begin{align}
 (\d{x}) = r_1 + \dots + r_{2g+2} - 2s_1 - 2s_2.
\end{align}
It follows that
\begin{align}
 \left(\frac{x^i \d{x}}{y}\right) = (g-i-1)s_1 + (g-i-1)s_2 + i t_1 + i t_2. 
\end{align}
Hence for $0\le i \le g-1$ the 1-forms $x^i \d{x}/y$ are holomorphic. Therefore the space of holomorphic 1-forms on $\Sigma$ is 
\begin{align}
 \Hc^0(\Sigma, K_{\Sigma}) = \left\{ \sum_{i=0}^{g-1} a_i \frac{x^i \d{x}}{y} \colon a_0,\dots, a_{g-1}\in\CC \right\}, \label{eq:hyperelliptic:holomorphic:forms}
\end{align}
where we have implicitly used that $\dim{\Hc^0(\Sigma, K_{\Sigma})} = g$. In our derivation of \eqref{eq:hyperelliptic:holomorphic:forms} we followed section III.7 in \cite{Farkas:Kra}.

Local coordinates on $\Sigma$ are obtained as follows: away from the ramification points $r_1, \dots,r_{2g+2}$ every value of $x$ determines two points on $\Sigma$,
\begin{align}
 \left(x, \sqrt{F(x)}\right), \quad \left(x, -\sqrt{F(x)}\right).
\end{align}
Thus the values of the function $x$, together with a choice of either the positive or the negative square root of $F(x)$, define a local coordinate. In the neighbourhood of a ramification point of $\pi$ the values of the function $y$ are used to define a local coordinate. For this to make sense the polynomial $F$ must be locally invertible at the ramification points of $\pi$, i.e.~$F'(x)\ne 0$ must hold whenever $F(x)=0$.

The hyperelliptic involution of $\Sigma$ is given by
\begin{align}
 &\sigma \colon \Sigma \to \Sigma, \\
 &\sigma(x,y) = (x,-y).
\end{align}
The $r_1, \dots,r_{2g+2}$ are precisely the fixed points of $\sigma$. The hyperelliptic involution extends to an automorphism of $\Sym^d(\Sigma)$,
\begin{align}
 \sigma \colon \Sym^d(\Sigma) &\to \Sym^d(\Sigma), \\
 \sum_i a_i p_i &\mapsto \sum_i a_i \sigma(p_i), \quad p_i\in\Sigma, \: a_i\in\ZZ, \: a_i>0, 
\end{align}
where once again we regard elements of $\Sym^d(\Sigma)$ as positive divisors. 

As stated in Theorem \ref{thm:geodesic}, we assume that $\Sigma$ is equipped with a metric such that $\sigma$ is an isometry. Note that such a metric always exists:\footnote{I thank Nick Manton for this observation.} if $h$ is any Riemannian metric on $\Sigma$, then the average $\half(h + \sigma^*h)$, where $\sigma^*$ denotes the pull-back under the hyperelliptic involution, is clearly invariant. Alternatively, using the uniqueness of the hyperbolic metric on $\Sigma$, one concludes that it is also necessarily invariant under $\sigma$. 

The action of $\sigma$ on $\Sigma$ pulls back to actions on the section $\phi$ and the covariant derivative $\covd = \d + A$. Since the Bogomolny equations \eqref{eq:ab:Bog:1}, \eqref{eq:ab:Bog:2} transform covariantly under these actions, $\sigma$ yields a diffeomorphism of the moduli space $\Sym^d(\Sigma)$. Moreover, the kinetic energy \eqref{eq:kinetic:energy} is invariant under the pulled back actions. Hence $\sigma$ yields in fact an isometry of the moduli space. Now it follows from Lemma 4.2 in \cite{Romao:isometries} that the fixed point set of $\sigma \colon \Sym^d(\Sigma) \to \Sym^d(\Sigma)$, denoted as $\fix_d(\sigma)$, is a geodesic submanifold of the moduli space. In the following we decompose $\fix_d(\sigma)$ into connected components, all of which are hence geodesic submanifolds of the moduli space.

\subsection{Linear systems as geodesic submanifolds of the moduli space}

The fixed point set of $\sigma \colon \Sym^d(\Sigma) \to \Sym^d(\Sigma)$ can be described in terms of the ramification points $r_1,\dots,r_{2g+2}$ as follows,
\begin{align}
 \fix_d(\sigma) = \left\{ D' + \sigma(D') + \sum_{i=1}^{2g+2} b_ir_i \colon \right. & D'\in\Sym^k(\Sigma), \: b_i\in\{0,1\},  \nonumber \\
												    & \left. 2k + \sum_{i=1}^{2g+2} b_i = d \right\}.
\end{align}
We introduce the sets
\begin{align}
 {\mathfrak D}_{d,b_1\cdots b_{2g+2}} = \left\{ D' + \sigma(D') + \sum_{i=1}^{2g+2} b_ir_i \colon D'\in\Sym^k(\Sigma), \, 2k + \sum_{i=1}^{2g+2} b_i = d \right\},
\end{align}
where a choice of $(b_1,\dots,b_{2g+2})\in\{0,1\}^{2g+2}$ has now been made. Note that ${\mathfrak D}_{d,b_1\cdots b_{2g+2}}$ is the empty set unless $\half \left( d-\sum b_i \right) \in \NN$. That a number of the ${\mathfrak D}_{d,b_1\cdots b_{2g+2}}$ are empty is a consequence of the fact that
\begin{align}
 2k + \sum_{i=1}^{2g+2} b_i = d
\end{align}
has exactly ${ 2g+2 \choose d-2k }$ solutions for fixed $d$ and $k$, while there are generally $2^{2g+2}$ different ways of choosing $(b_1,\dots,b_{2g+2})\in\{0,1\}^{2g+2}$. 

Different ${\mathfrak D}_{d,b_1\cdots b_{2g+2}}$ are clearly disjoint, i.e.~$(b_1,\dots,b_{2g+2}) \ne (b_1',\dots,b_{2g+2}')$ implies
\begin{align}
 {\mathfrak D}_{d,b_1\cdots b_{2g+2}} \cap {\mathfrak D}_{d,b_1'\cdots b_{2g+2}'} = \emptyset.
\end{align}
We thus have the decomposition, 
\begin{align}
 \fix_d(\sigma) = \bigcup_{\substack{{b_i\in\{0,1\},}\\{\half \left( d-\sum b_i \right) \in \NN}}} {\mathfrak D}_{d,b_1\cdots b_{2g+2}}, \label{eq:fix:decomposition}
\end{align}
which we will now show is a decomposition of $\fix_d(\sigma)$ into connected components.

Note that for any $D_1,D_2 \in \Sym^k(\Sigma)$ the divisors $D_1 + \sigma(D_1)$ and $D_2 + \sigma(D_2)$ are linearly equivalent. This is because there always exists a rational function $f_{D_2-D_1} \colon \CP{1} \to \CP{1}$ of degree $k$ such that
\begin{align}
 D_2 - D_1 + \sigma(D_2) - \sigma(D_1) = ( f_{D_2-D_1} \circ \pi ).
\end{align}
It follows that the non-empty ${\mathfrak D}_{d,b_1\cdots b_{2g+2}}$ are linear systems. Each of these linear systems sits inside a complete linear system,
\begin{align}
 {\mathfrak D}_{d,b_1\cdots b_{2g+2}} \subset \PP{\mathfrak L}(D_{b_1\cdots b_{2g+2}}),
\end{align}
where
\begin{align}
 D_{b_1\cdots b_{2g+2}} = D' + \sigma(D') + \sum_{i=1}^{2g+2} b_ir_i,
\end{align}
for an arbitrary $D'\in\Sym^k(\Sigma)$. It also follows that ${\mathfrak D}_{d,b_1\cdots b_{2g+2}} \cong \CP{k}$, thus establishing the connectedness of the non-empty ${\mathfrak D}_{d,b_1\cdots b_{2g+2}}$.

Based on the decomposition \eqref{eq:fix:decomposition} we can now state the Lemma from which Theorem \ref{thm:geodesic} will follow.

\newtheorem{geodesic:decomposition}[near:Bradlow:limit]{Lemma}
\begin{geodesic:decomposition} \label{thm:geodesic:decomposition}
Let $\Sigma$ be a hyperelliptic Riemann surface, equipped with a metric such that the hyperelliptic involution of $\Sigma$ is an isometry, and let $d$ denote the vortex number. Every non-empty linear system ${\mathfrak D}_{d,b_1\cdots b_{2g+2}}$ is a geodesic submanifold of the vortex moduli space $\Sym^d(\Sigma)$. 
\end{geodesic:decomposition}

Before turning to the proof of Theorem \ref{thm:geodesic}, we conclude this section with a few remarks on how the ${\mathfrak D}_{d,b_1\cdots b_{2g+2}}$ embed into the corresponding complete linear systems, i.e.~into fibres of the Abel--Jacobi map. Crucial to our observations is the following proposition, a proof of which we give in appendix \ref{app:linear:equivalence}.

\newtheorem{inequivalent:divisors}{Proposition}
\begin{inequivalent:divisors} \label{thm:inequivalent:divisors}
Let $\Sigma$ be a hyperelliptic Riemann surface. Let $D_{b_1\cdots b_{2g+2}}$ and $D_{b_1'\cdots b_{2g+2}'}$ be divisors of degree $d\in\NN$ such that
\begin{align}
 &D_{b_1\cdots b_{2g+2}} = D_1 + \sigma(D_1) + \sum_{i=1}^{2g+2} b_ir_i, \\
 &D_{b_1'\cdots b_{2g+2}'}  = D_2 + \sigma(D_2) + \sum_{i=1}^{2g+2} b_i'r_i,
\end{align}
where $b_i, b_i' \in\{0,1\}$ and
\begin{align}
 &D_1 \in \Sym^k(\Sigma), \quad k = \half \left( d - \sum_{i=1}^{2g+2}  b_i \right) \in \NN, \\
 &D_2 \in \Sym^{k'}(\Sigma), \quad k' = \half \left( d - \sum_{i=1}^{2g+2}  b_i' \right) \in \NN.
\end{align}
The following hold:
\begin{enumerate}
 \item If $(b_1,\dots,b_{2g+2}) \ne (b_1',\dots,b_{2g+2}')$ and, after canceling common terms, $\sum (b_i + b_i') \ne 2g + 2$, then $D_{b_1\cdots b_{2g+2}}$ and $D_{b_1'\cdots b_{2g+2}'}$ are linearly inequivalent.
 \item If $(b_1,\dots,b_{2g+2}) = (b_1',\dots,b_{2g+2}')$ or, after canceling common terms, $\sum (b_i + b_i') = 2g + 2$, then $D_{b_1\cdots b_{2g+2}}$ and $D_{b_1'\cdots b_{2g+2}'}$ are linearly equivalent.
\end{enumerate}
\end{inequivalent:divisors}
 
It follows that for $d < g+1$ all of the non-empty ${\mathfrak D}_{d,b_1\cdots b_{2g+2}}$ are subsets of distinct fibres of $\AJ$. For $d = g+1$ this is no longer true as the second statement in the Proposition applies to
\begin{align}
 &b_1 = \dots = b_{g+1} = 1, \quad b_{g+2}  = \dots = b_{2g+2} = 0, \\
 &b_1' = \dots = b_{g+1}' = 0, \quad b_{g+2} ' = \dots = b_{2g+2}' = 1.
\end{align} 
It is worthwhile checking that for $d \ge g+1$ and $\sum (b_i + b_i') = 2g + 2$ there is enough room, so to speak, in the fibres of $\AJ$ to contain the disjoint sets ${\mathfrak D}_{d,b_1\cdots b_{2g+2}}$ and ${\mathfrak D}_{d,b_1'\cdots b_{2g+2}'}$. Note that
\begin{align}
 &{\mathfrak D}_{d,b_1\cdots b_{2g+2}} \cong \CP{k}, \\
 &{\mathfrak D}_{d,b_1'\cdots b_{2g+2}'} \cong \CP{k'},
\end{align}
where $k$ and $k'$ are as in the Proposition. It follows from the Riemann--Roch Theorem that
\begin{align}
 \dim{\PP\Hc^0(\Sigma, \Ob{D_{b_1\cdots b_{2g+2}}})} \ge d - g. 
\end{align}
Since
\begin{align}
 k + k' = d - g - 1,
\end{align}
it is indeed possible for $\CP{k}$ and $\CP{k'}$ to embed into $\PP\Hc^0(\Sigma, \Ob{D_{b_1\cdots b_{2g+2}}})$ without intersecting.

Based on Lemma \ref{thm:geodesic:decomposition} it is easy to establish Theorem \ref{thm:geodesic}. We do so in the following sections, each of which is dedicated to one of the fibres of $\AJ$ that features in Theorem \ref{thm:geodesic}. We also comment on why these fibres are of particular interest.

\subsection{Holomorphic 1-forms as Higgs fields}

Let the vortex number be $d = 2g-2$, and let ${\mathfrak p} = \AJ(K_{\Sigma})$. Because of the identification $\AJ^{-1}\{{\mathfrak p}\} \cong \PP\Hc^0(\Sigma, K_{\Sigma})$, the submanifold $\AJ^{-1}\{{\mathfrak p}\}$ corresponds to vortex configurations where the Higgs field $\phi$ is a holomorphic 1-form. This is an interesting special case since, by virtue of \eqref{eq:hyperelliptic:holomorphic:forms}, one can give explicit formulae for $\phi$. One can also give expressions analogous to the ones in \eqref{eq:hyperelliptic:holomorphic:forms} for the holomorphic 1-forms on non-hyperelliptic surfaces, see e.g.~\cite{Griffiths:curves}. Furthermore there may be a relationship to Higgs bundles as discussed in the mathematical literature, cf.~\cite{Hitchin:self:duality}, which we do not explore in this paper.

We claim that ${\mathfrak D}_{2g-2,0\cdots0} \cong \PP\Hc^0(\Sigma, K_{\Sigma})$. To see this, first note that the meromorphic function $x$ satisfies $x\circ\sigma = x$. This implies
\begin{align}
 &s_2 = \sigma(s_1).
\end{align}
Therefore the divisor of $\omega_0 = \d{x}/y \in \Hc^0(\Sigma, K_{\Sigma})$ is
\begin{align}
 \left(\frac{\d{x}}{y}\right) = (g-1)s_1 + (g-1)\sigma(s_1),
\end{align}
and hence $(\omega_0) \in {\mathfrak D}_{2g-2,0\cdots0}$. Since $\PP{\mathfrak L}((\omega_0))$ is the complete linear system associated to the divisor $(\omega_0)$, it follows that
\begin{align}
 {\mathfrak D}_{2g-2,0\cdots0} \subset \PP{\mathfrak L}((\omega_0)) \cong \PP\Hc^0(\Sigma, K_{\Sigma}). 
\end{align}
Noting that $\PP\Hc^0(\Sigma, K_{\Sigma}) \cong \CP{g-1}$ and also ${\mathfrak D}_{2g-2,0\cdots0} \cong \CP{g-1}$ proves our claim.

\subsection{Symmetric vortex pairs}

Now we consider configurations of two vortices on $\Sigma$ that are symmetric under the hyperelliptic involution. Such vortex configurations are described by divisors in the fixed point set
\begin{align}
 \fix_2(\sigma) = {\mathfrak D}_{2,0\cdots0} \cup \left\{ \sum_{i=1}^{2g+2} b_ir_i \colon b_i\in\{0,1\}, \sum_{i=1}^{2g+2} b_i = 2 \right\}.
\end{align}
We claim that ${\mathfrak D}_{2,0\cdots0}$ is a fibre of the Abel--Jacobi map. To see this, first note that $D=s_1+s_2\in{\mathfrak D}_{2,0\cdots0}$, and hence
\begin{align}
 \CP{1} \cong {\mathfrak D}_{2,0\cdots0} \subset \PP{\mathfrak L}(D).
\end{align}
The meromorphic function $x$ lies in ${\mathfrak L}(D)$. Let $h\colon\Sigma\to\CC$ be a non-constant meromorphic function such that $h\in{\mathfrak L}(D)$. Then there exists a constant $c\in\CC$ such that $h-cx$ has no pole at $s_1$, i.e.
\begin{align}
 (h-cx) \ge -s_2.
\end{align}
The only meromorphic functions on $\Sigma$ which have at most a single pole are the constant functions. Hence $h-cx$ is constant. Thus we can conclude that
\begin{align}
 {\mathfrak L}(D) = \spn_{\CC} \{1,x\},
\end{align}
and hence
\begin{align}
 \CP{1} \cong {\mathfrak D}_{2,0\cdots0} = \PP{\mathfrak L}(D) = \AJ^{-1}\{\AJ(D)\}. 
\end{align}
To conclude the proof of Theorem \ref{thm:geodesic}, it suffices to note that $D = \pi^{-1}(\pi(s_1))$. 

Pairs of vortices that are symmetric under $\sigma$ provide a simple but novel example of vortex scattering on a compact Riemann surface. We briefly elaborate this example. First note that $\fix_2(\sigma) \setminus {\mathfrak D}_{2,0\cdots0}$ is a discrete set. Hence moving vortex pairs that are symmetric under $\sigma$ at all times correspond to divisors in ${\mathfrak D}_{2,0\cdots0}$. With $D=s_1+s_2$, as before, we have that $\dim{\Hc^0(\Sigma, \Ob{D})}=2$. Therefore let 
\begin{align}
 \Hc^0(\Sigma, \Ob{D}) = \spn_{\CC}\{ \phi_1, \phi_2 \}.
\end{align}
We describe the motion of symmetric vortex pairs by the 1-parameter family of Higgs fields
\begin{align}
 \phi_t(p) = \lambda_1(t)\phi_1(p) + \lambda_2(t)\phi_2(p), \quad t\in\RR, \: p\in\Sigma, \: \lambda_1(t),\lambda_2(t)\in\CC,
\end{align}
where for all $t$ we assume that $\lambda_1(t),\lambda_2(t)$ are chosen such that $\phi_t$ satisfies the Bogomolny equations. At time $t$ the vortices are centred at the two zeros of $\phi_t$. More concretely, the vortices are located at $p_1(t), p_2(t)\in\Sigma$ such that 
\begin{align}
 f(p_{1,2}(t)) = - \frac{\lambda_1(t)}{\lambda_2(t)}, \label{eq:vortex:centres}
\end{align}
where $f$ is the meromorphic function
\begin{align}
 f = \frac{\phi_2}{\phi_1} \colon \Sigma \to \CC.
\end{align}
Regarded as a map from $\Sigma$ to $\CP{1}$, $f$ is 2-1 and has $2g+2$ ramification points. Since $\Sigma$ is hyperelliptic, we can apply the Theorem in subsection III.7.3 in \cite{Farkas:Kra} which states that $f$ and the projection map $\pi$ are related by a M\"obius transformation. In particular, $f$ and $\pi$ have the same ramification points $r_1,\dots,r_{2g+2}$. Vortices scatter when $p_1(t) = p_2(t)$, and because of \eqref{eq:vortex:centres} this can only happen at a ramification point. The $r_1,\dots,r_{2g+2}$ are of course characteristic of the hyperelliptic surface $\Sigma$. It is a new observation that vortex pairs which respect the hyperelliptic involution can only scatter at these points. 

We can also verify in our particular example the well-known fact that vortices scatter through right angles. To this end, let $p_1(0) = p_2(0)$ be a ramification point of $f$, in a neighbourhood of which we can define a local coordinate $w$ by
\begin{align}
 f(x) + \frac{\lambda_1(0)}{\lambda_2(0)} = w^2.
\end{align}
To obtain the vortex centres in the coordinate $w$, we use the relation
\begin{align}
 w(t)^2 = - \frac{\lambda_1(t)}{\lambda_2(t)} + \frac{\lambda_1(0)}{\lambda_2(0)} ,
\end{align}
and expand the right-hand side in $t$,
\begin{align}
 w(t)^2 = c \, t + \dots, \quad c = - \frac{\lambda_2(0) {\dot \lambda}_1(0) - \lambda_1(0) {\dot \lambda}_2(0)}{\lambda_2^2(0)}.
\end{align}
Therefore, for $t$ close to zero,
\begin{align}
 w(t) = \pm\sqrt{c} \sqrt{t},
\end{align}
and the phase of $w$ changes by $\frac{\pi}{2}$ as $t$ passes through zero. This means that the vortex centres, located at time $t$ at the two possible values of $w(t)$, scatter at a right angle in a head-on collision. 

To conclude this paper, we remark on the generality of our method of obtaining geodesic submanifolds of the vortex moduli space: the statement and proof of Lemma \ref{thm:geodesic:decomposition} do not depend on being near the Bradlow limit. However, close to the Bradlow limit the linear systems ${\mathfrak D}_{d,b_0\cdots b_{2g+2}}$ are subject to Theorem \ref{thm:FS:metric}. This is because restricting the Fubini--Study metric to an embedded projective space yields the Fubini--Study metric on this space. It would be interesting to understand better the restrictions of the moduli space metric to linear systems away from the Bradlow limit.

\section{Acknowledgements}

This work was initiated and Theorems \ref{thm:FS:metric} and \ref{thm:geodesic} were obtained during my PhD research. I wish to thank my research supervisor, Nick Manton, for numerous helpful discussions and for comments on the manuscript of this paper. I acknowledge a discussion with John Ottem on the Abel--Jacobi map, and I thank Martin Wolf and Moritz H\"ogner for discussions on holomorphic structures on vector bundles. This work was financially supported by the Cambridge Philosophical Society.

\appendix
\numberwithin{equation}{section}

\section{Linearly equivalent and inequivalent divisors on hyperelliptic surfaces} \label{app:linear:equivalence}

In this appendix we give a proof of Proposition \ref{thm:inequivalent:divisors}. As a preparation, note that after renumbering the ramification points $r_1,\dots,r_{2g+2}$, if necessary, we can write
\begin{align}
 &\sum_{i=1}^{2g+2} b_i r_i  =       r_1 + \dots + r_l  \ph{ +   r_{l+1} + \dots + r_{l+m}} + r_{l+m+1} + \dots +  r_{l+m+n}, \\
 &\sum_{i=1}^{2g+2} b_i' r_i = \ph{r_1 + \dots + r_l         + } r_{l+1} + \dots + r_{l+m}  + r_{l+m+1} + \dots +  r_{l+m+n},
\end{align}
where $l+m+n \le 2g+2$. Note in particular that $l+m = 2g+2$ is possible only if $n=0$. The notation
\begin{align}
 &x_i = x(r_i), \quad\text{for } i = 1, \dots, 2g+2,
\end{align}
will also be useful in proving the proposition.

We first verify the second statement of Propositon \ref{thm:inequivalent:divisors}. If $(b_1,\dots,b_{2g+2}) = (b_1',\dots,b_{2g+2}')$, then all that needs to be shown is that the divisors $D_1+\sigma(D_1)$ and $D_2+\sigma(D_2)$ are linearly equivalent. This was done as part of the derivation of Lemma \ref{thm:geodesic:decomposition}. If, on the other hand, after canceling common terms, $\sum (b_i + b_i') = 2g + 2$, then $l+m=2g+2$. It follows from
\begin{align}
 2k + l = 2k' + m
\end{align}
that $m-l$ is even. Recalling that $s_2 = \sigma(s_1)$, it now suffices to show that the divisor
\begin{align}
 r_1 + \dots + r_{l} - r_{l+1} - \dots - r_{2g+2} + \frac{m-l}{2}s_1 + \frac{m-l}{2}s_2
\end{align}
is principal. From \eqref{eq:x:divisor} we obtain the following principal divisors,
\begin{align}
 (x-x_i) = 2r_i - s_1 - s_2, \quad\text{for } i = 1, \dots, 2g+2.
\end{align}
Together with \eqref{eq:y:divisor} this leads to 
\begin{align}
 \left( \frac{y}{(x-x_{l+1}) \cdots (x-x_{2g+2})}\right) &= r_1 + \dots + r_{l} - r_{l+1} - \dots - r_{2g+2} \nonumber \\
 									   &\ph{=} - (g+1-m)s_1 - (g+1 - m)s_2,
\end{align}
as desired.

We now prove the first statement of the proposition by showing that the linear equivalence of $D_{b_1\cdots b_{2g+2}}$ and $D_{b_1'\cdots b_{2g+2}'}$ leads to a contradiction. Let us therefore assume that there exists a meromorphic function $h \colon \Sigma \to \CC$ with divisor
\begin{align}
 (h) = {\tilde D} + \sigma({\tilde D}) + r_1 + \dots + r_l - r_{l+1} - \dots - r_{l+m}, \label{eq:divisor:h}
\end{align}
where
\begin{align}
 {\tilde D} \in \Sym^{\tilde k}(\Sigma), \quad {\tilde k} = \half \sum_{i=1}^{2g+2} (b_i'-b_i) = \frac{m-l}{2} \in \NN.
\end{align}
Note that we have also assumed, without loss of generality, that $m \ge l$. From $(b_1,\dots,b_{2g+2}) \ne (b_1',\dots,b_{2g+2}')$ and $\sum (b_i + b_i') \ne 2g + 2$, after canceling common terms, it follows that 
\begin{align}
 1 \le l+m < 2g+2. \label{eq:rpoints:bounds}
 \end{align}
Now let ${\tilde D} = p_1 + \dots + p_{\tilde k}$, and introduce the notation
\begin{align}
 z_i = x(p_i), \quad\text{for } i = 1, \dots, {\tilde k}.
\end{align}
After multiplying by an overall constant factor, if necessary, we can assume that the meromorphic function $h^2$ is given by
\begin{align}
  h^2 &= (x-x_1) \dots (x-x_l) \nonumber \\
      &\ph{=} \cdot (x-x_{l+1})^{-1} \dots (x-x_{l+m})^{-1} \nonumber \\
      &\ph{=} \cdot (x-z_1)^2 \dots (x-z_{\tilde k})^2. 
\end{align}
This expression is invariant under the hyperelliptic involution, i.e.~$h(\sigma(p))^2 = h(p)^2$ for every $p\in\Sigma$. We therefore have $h(\sigma(p)) = \pm h(p)$ for every $p\in\Sigma$. Since the ramification points $r_1, \dots, r_{l+m}$ appear with multiplicity 1 in \eqref{eq:divisor:h}, the function $h$ does not descend to a meromorphic function on $\CP{1}$. Hence there exists at least one $p\in\Sigma$ for which $h(\sigma(p)) \ne h(p) \ne 0$. Since $h$ is smooth and vanishes only at isolated points, it follows that $h\circ\sigma = -h$.

The meromorphic function $f \colon \Sigma \to \CC$, defined by
\begin{align}
 f &= (x-x_{l+1}) \dots (x-x_{l+m}) \nonumber \\
   &\ph{=} \cdot (x-z_1)^{-1} \dots (x-z_{\tilde k})^{-1} \nonumber \\ 
   &\ph{=} \cdot h,
\end{align}
satisfies the equation
\begin{align}
 f^2 = (x-x_1) \dots (x-x_l) (x-x_{l+1}) \dots (x-x_{l+m}).
\end{align}
Therefore the pairs $(x, f)$ define a hyperelliptic Riemann surface $\Sigma'$ of genus $g' = \frac{l+m -2}{2}$. From \eqref{eq:rpoints:bounds} it follows that $g' < g$. Recall from the definition of $\Sigma$ that its points are given as pairs $(x,y)$. Since $h$ does not descend to a function on $\CP{1}$, it cannot depend only on $x$, but must also depend on $y$. It follows that $f$ too depends on $y$. The following map makes the $y$ dependence of $f$ explicit,
\begin{align}
 \Sigma \ni (x,y)  &\mapsto (x,f(x,y)) \in \Sigma'.
\end{align}
Note that this is a non-constant holomorphic map between compact Riemann surfaces, and hence is onto. To see that it is also 1-1, assume that there exist $(x,y),(x,-y)\in\Sigma$ such that
\begin{align}
 (x,f(x,y)) = (x,f(x,-y)).
\end{align}
This implies that $h(x,y) = h(x,-y)$, contradicting $h\circ\sigma = -h$.

From the meromorphic function $h$ we have thus constructed a biholomorphic  map between $\Sigma$ and $\Sigma'$. Since these surfaces have different genera, they cannot even be homeomorphic as smooth manifolds. This is the desired contradiction which establishes the first statement in Propositon \ref{thm:inequivalent:divisors}.

\bibliographystyle{utphys-new}  
\clearpage 
\addcontentsline{toc}{section}{References}
\bibliography{references}

\end{document}